# Zonal Functions on SO(p,q) Groups

## Bala Ali RAJABOV

*Institute of Physics, Azerbaijan National Academy of Sciences,*

*Javid ave. 33, Baku Azerbaijan*

**Abstract**

Explicit expressions for zonal spherical functions of SO(p,q) matrix groups are obtained using a generalized hypergeometric series of two variables.

### 1. Introduction

The SO(p,q) groups of pseudo-orthogonal matrices and their representations are broadly used in different fields of physics, particularly in quantum field theory, high-energy physics, cosmology and solid-state physics [1-2].

Matrix elements of irreducible unitary representations (IUR) of these groups, particularly spherical functions, play an essential role in the theory of representations of SO(p,q) groups. These functions in the case of SO(p, 1) groups are investigated in details [3-4]. Unlike them, the construction of spherical functions of SO(p,q), $p \geq q \geq 2$ groups has not yet been completed. N. Ya. Vilenkin's and A.P. Pavluk's publications [2], in which the "relationship between the spherical functions of matrix groups and Herz's functions of matrices argument was established [5], have been receiving remarkable attention.

The purpose of this paper is to find the spherical functions of SO(p,q) groups when $p \geq q \geq 2$. The main result consists of the fact that zonal spherical functions of the groups SO(p,q), $p \geq q \geq 2$, are expressed as generalized hypergeometric Horn's functions of two variables by a unique formula, in spite of essentially different forms of the integral representations of spherical functions at $p \geq q \geq 3$ and $p \geq 2, q = 2$. The preliminary statements about the obtained results were made in a Report of the Academy of Sciences of Azerbaijan Republic [6-7].
Furthermore, the obtained formula has been proved to be valid for $q = 1$, i.e., for SO(p,1), $p \geq 2$ groups.

This work is prolongation of the publication [12]. Unfortunately, at printing paper [12] some errors were supposed. In this publication we reduce rectified expressions for some formulas [12], and also new expressions for zonal functions.

### 2. Most degenerate irreducible unitary representations of SO(p,q) groups

The most degenerate representations of groups SO(p,q), i.e., connected components of units of groups of motions of $(p + q)$-dimensional vector space holding invariant quadratic form:

$$[k,k] = k_1^2 + k_2^2 + \cdots + k_q^2 - k_{q+1}^2 - \cdots - k_{p+q}^2,$$

are given by a complex number, σ, and a number, ε, which take the values 0 and 1, and are constructed in the space of homogeneous functions, F(.), of given parity and defined on the cone *[k, k] = 0*, [1 — 6] .

Let us denote by $D^{(\sigma,\varepsilon)}$ the space of infinitely differentiable functions, F(.) , denned on the cone $[k,k]=0$ without the point $k = 0$ and satisfying the following condition:

$$F(ak) = |a|^{\sigma} F(k) sign^{\varepsilon}(a), \quad a \neq 0, \quad \varepsilon = 0, 1,$$

where *a* is any real number:

The action of the operator $T^{(\sigma,\varepsilon)}(g)$ of SO(p,q) in the space $D^{(\sigma,\varepsilon)}$ is determined as follows:

$$T^{(\sigma,\varepsilon)}(g)F(k) = F(g^{-1}k), \quad g \in SO(p,q). \tag{1}$$

The space of the representation and the representation itself may have different realizations. Let us consider one of them.

Let us introduce the spherical system of coordinates on the cone *[k, k] = 0*:

$$k = \omega(\cos\phi, \eta\sin\phi, \cos\chi, \xi\sin\chi) \tag{2}$$

where $0 < \omega < \infty$, and η and ξ are *(q — 1)* -dimensional and *(p — 1)* -dimensional unit vectors, respectively.

Spherical angles φ and χ change within the following limits:
$0 \leq \phi \leq \pi, \quad 0 \leq \chi \leq \pi$, if $p \geq q \geq 3$;
$0 \leq \phi < 2\pi, \quad 0 \leq \chi \leq \pi$, if $p \geq 2, q = 2$ (in this case η is reduced to a constant value which we take as η = 1);
$0 \leq \phi < 2\pi, \quad 0 \leq \chi < 2\pi,$, if $p = q = 2$ (in this case η and ξ are reduced to a constant values which we take as η = ξ = 1 ) .

Let us consider the reduction of functions from $D^{(\sigma,\varepsilon)}$ on the cross-section, ω = 1 , of the cone *[k,k] = 0*:

$$f(\phi,\eta;\chi,\xi) = F(k)\big|_{\omega=1}. \tag{3}$$

Then, in virtue of homogeneity of F(.) we obtain:

$$F(k) = \omega^{\sigma} f(\phi,\eta;\chi,\xi) \tag{4}$$

From equations (3)-(4) it is evident that function f(.) also has the given parity ε:

$$f(\phi,\eta;\chi,\xi) = (-1)^{\varepsilon} f(\pi - \phi, -\eta; \pi - \chi, -\xi)$$

The mappings (3)-(4) establish a one-to-one correspondence between $D^{(\sigma,\varepsilon)}$ and the space of infinitely differentiable functions defined on $S^q \otimes S^p$ . From Eq. (1)-(4) we obtain the following expression, using the same notation for the space and the operator ($D^{(\sigma,\varepsilon)}$ and $T^{(\sigma,\varepsilon)}$, respectively) of the representations:

$$T^{(\sigma,\varepsilon)}(g)f(\phi,\eta;\chi,\xi) = (\omega_g/\omega)^{\sigma} f(\phi_g,\eta_g;\chi_g,\xi_g), \tag{5}$$

where $\omega_g, \phi_g, \eta_g, \chi_g, \xi_g$ are found from the relation of: $g^{-1}k = k_g$. In the

space $D^{(\sigma,\varepsilon)}$, let us introduce the scalar product[1]:

$$(f_1, f_2) = \int f_1(\phi,\eta;\chi,\xi)\overline{f_2(\phi,\eta;\chi,\xi)}(\sin\phi)^{q-2}d\phi(\sin\chi)^{p-2}d\chi(d\eta)(d\xi), \quad (6)$$

where $(d\eta)$ and $(d\xi)$ are normalized measures on $S^{q-1}$ and $S^{p-1}$, respectively, [3].

In the case q=2 or p=q=2, the differentials $(d\eta)$ or $(d\eta)(d\xi)$ are omitted because they are constants, and the other variables are integrated over the whole region.

It follows directly form Eq. (5)-(6) that the scalar product (6) is invariant for $\operatorname{Re}\sigma = -\dfrac{p+q-2}{2}$, $\varepsilon = 0,1$. By filling the space $D^{(\sigma,\varepsilon)}$ with the scalar product (6), we obtain the Hilbert space, $H^{(\sigma,\varepsilon)}$, and the most degenerated IUR's of the continuous principal series of the group $SO(p, q)$. It is easy to show that $(\sigma, \varepsilon)$ and $(2 - p - q - \sigma, \varepsilon)$ representations are unitarily equivalent [6].

Formulas (5)-(6) allow the integral representations for matrix elements of the operator of hyperbolic rotations on the surface $(\kappa_1, \kappa_{n+1})$ in canonical basis to be determined [3]. With respect to the canonical basis vector (which is invariant under the subgroup $S^q \otimes S^p$ the matrix elements which are the zonal spherical functions of group $SO(p,q)$ as determined in [3] are of interest. Denoting these functions by $Z_\sigma^{[p,q]}(\alpha)$ we obtain from Eq. (5)-(6) the following integral representations:

1. In the case of $p \geq q \geq 3$:

$$Z_\sigma^{[p,q]}(\alpha) = \frac{\Gamma(p/2)\Gamma(q/2)}{\pi\Gamma\left[\dfrac{p-1}{2}\right]\Gamma\left[\dfrac{q-1}{2}\right]} \int_{-1}^{+1}\int_{-1}^{+1} \Theta^{\sigma/2}(1-x^2)^{\frac{p-3}{2}}(1-y^2)^{\frac{q-3}{2}} dxdy; \quad (7)$$

2. In the case of $p \geq 3$, $q = 2$:

$$Z_\sigma^{[p,2]}(\alpha) = \frac{\Gamma(p/2)}{\pi^{3/2}\Gamma\left[\dfrac{p-1}{2}\right]} \int_0^{2\pi}\int_{-1}^{+1} \Theta^{\sigma/2}(1-x^2)^{\frac{p-2}{2}} d\phi\, dx; \quad (8)$$

3. In the case of p=q=2:

$$Z_\sigma^{[3,2]}(\alpha) = \frac{1}{4\pi^2}\int_0^{2\pi}\int_0^{2\pi}\Theta^{\sigma/2} d\phi d\chi. \quad (9)$$

In (7)-(9) we used the following notation:

$$\Theta = (\cos\phi\, ch\alpha - \cos\chi\, sh\alpha)^2 + \sin^2\phi = 1 + (x^2 + y^2)sh^2\alpha - 2xy\, sh\alpha\, ch\alpha, \quad (10)$$

$$x = \cos\chi, \quad y = \cos\phi.$$

It is to be noted that zonal function exists only for even representations ($\varepsilon = 0$).

---
[1] The bar denotes complex conjugation.

## 3. Zonal functions and Horn's functions

The main formulas used for the calculation of zonal functions of SO(p,q) groups are the integral representations (7)-(9) and the Taylor expansions for the function $\Theta^{\sigma/2}$ given as:

$$\Theta^{\sigma/2} = \sum_{v=0}^{1} \frac{(-\sigma \, xy \, th\alpha)^v}{ch\alpha} \sum_{l=0}^{\infty} \frac{(v+1/2)_l}{l!} th^{2l}\alpha \, F_2\left(v-\frac{\sigma}{2},-l,-l;v+\frac{1}{2},v+\frac{1}{2};x^2,y^2\right) \quad (11)$$

Here we use the following notation for Pochhammer's symbols, $(a)_n$, and Appell's functions of a second kind [8]:

$$(a)_n = \Gamma(a+n)/\Gamma(a),$$

$$F_2\left(v-\frac{\sigma}{2},-l,-l;v+\frac{1}{2},v+\frac{1}{2};x^2,y^2\right) = \sum_{m,n=0}^{l} \frac{(v-\sigma/2)_{m+n}(-l)_m(-l)_n}{(v+1/2)_m(v+1/2)_n \, m! n!} x^{2m} y^{2n}$$

Series (11) converges uniformly and absolutely for sufficiently small values of $\alpha$, namely at $ch2\alpha < 3$ (the sufficient condition !).

Substituting the expansion into Eq.(11), alternatively in (7)-(9), and integrating for zonal function of SO(p,q), $p \geq q \geq 2$, groups, we obtain the following expression:

$$Z_\sigma^{[p,q]}(\alpha) = \sum_{m=0}^{\infty} \frac{\left(\frac{1}{2}\right)_m \left(-\frac{\sigma}{2}\right)_m \left(1-\frac{\sigma+q}{2}\right)_m}{(p/2)_m (q/2)_m \, m! \, ch\alpha} th^{2m}\alpha \; {}_3F_2\left(\begin{array}{c} -m, \; 1-m-\frac{p}{2}, \; \frac{\sigma+q}{2}; \\ 1-m+\frac{\sigma}{2}, \; \frac{\sigma+q}{2}-m; \end{array} 1\right) \quad (12)$$

Furthermore, assuming $q = 1$ in (12), after necessary simplifications, we obtain the following formula:

$$Z_\sigma^{[p,1]}(\alpha) = ch^\sigma\alpha \; {}_2F_1\left[-\frac{\sigma}{2}, \; \frac{1-\sigma}{2}; \; \frac{p}{2}; \; th^2\alpha\right],$$

which exactly coincides with the expression for zonal functions of *SO(p, 1)* groups. Thus, formula (12) is valid for all $SO(p,q)$, $p \geq 2$, $q \geq 1$ groups. Expression (12) for zonal functions of *SO(p, q)* groups can be rewritten more compactly with generalized hypergeometric functions of two variables, (A1)-(A2):

$$Z_\sigma^{[p,q]}(\alpha) = \frac{1}{ch\alpha} \; {}_4F_2\left[\begin{array}{cc|cc} 10 & -\sigma/2; & 1-(\sigma+q)/2 \\ 01 & (\sigma+q)/2 & \\ 11 & 1/2 & \\ \hline 11 & q/2 & \\ 10 & p/2 & \end{array} \; th^2\alpha, th^2\alpha \right] \quad (13)$$

From the results of Horn's theory for hypergeometric series of two variables, it follows that the series (13) converges for all finite $\alpha$.

Rearranging indexes of toting it is possible to receive alternate expressions for zonal functions:

$$Z_\sigma^{[p,q]}(\alpha) = \frac{1}{ch\alpha}\ _4F_2 \begin{bmatrix} 10 & -\sigma/2; & 1-(\sigma+p)/2 \\ 01 & (\sigma+p)/2 & \\ 11 & 1/2 & \\ 11 & p/2 & \\ 10 & q/2 & \end{bmatrix} th^2\alpha, th^2\alpha \quad (14)$$

The formulas (13) - (14) express properties of a symmetry of zonal functions concerning permutation p and q.

The above-presented results can be summarized in terms of the following theorem:

*Theorem:* Zonal spherical functions for all $SO(p,q)$, $p \geq 2$, $q \geq 1$ groups in general are expressed by hypergeometric functions of two variables according to the formulas (12)-(13).

In order to complete the proof of the theorem it is sufficient to note that the zonal functions (7)-(9) are analytic functions of *a* and sufficient to use the principle of monodromy.

## 6. Appendix

The generalized hypergeometric Horn's series with *r* variables is defined as follows [10]:

$$\sum_{n_1\cdots n_r=0}^{\infty} \frac{\prod_{\alpha=1}^{p}\left(a_\alpha, \sum_{j=1}^{r} u_{\alpha j} n_j\right)}{\prod_{\beta=1}^{q}\left(b_\beta, \sum_{j=1}^{r} v_{\beta j} n_j\right)} \cdot \prod_{i=1}^{r} \frac{X_i^{n_i}}{n_i!}, \quad (A1)$$

$$\sum_\alpha^p u_{\alpha j} = \sum_\beta^q v_{\beta j} + 1, \quad 1 \leq j \leq r,$$

where

$$(\lambda, n) = \frac{\Gamma(\lambda+n)}{\Gamma(\lambda)} = (\lambda)_n, \quad (\lambda, 0) = 1.$$

It is convenient to denote the series (A1) by the notations [11]:

$$_pF_q\begin{bmatrix} u_{\alpha_1 1}, & \cdots & u_{\alpha_1 r} & \{a_{\alpha_1}\} \\ \cdots & \cdots & \cdots & \cdot \\ \cdots & \cdots & \cdots & \cdot \\ \cdots & \cdots & \cdots & \cdot \\ u_{\alpha_s 1}, & \cdots & u_{\alpha_s r} & \{a_{\alpha_s}\} \\ \hline v_{\beta_1 1}, & \cdots & v_{\beta_1 r} & \{b_{\beta_1}\} \\ \cdots & \cdots & \cdots & \cdot \\ \cdots & \cdots & \cdots & \cdot \\ \cdots & \cdots & \cdots & \cdot \\ v_{\beta_t 1}, & \cdots & v_{\beta_t r} & \{b_{\beta_t}\} \end{bmatrix} X_1, \cdots X_r \Bigg], \qquad (A2)$$

$$s \leq p, \quad r \leq q,$$

where the multiples in the numerator (denominator) of (A2), corresponding to same values of $u_{\alpha j}$ $(v_{\beta j})$, $1 \leq j \leq r$, are unified in the same row of (A2), i.e., the set of the same values of $\alpha, 1 \leq \alpha \leq p$ $(\beta, 1 \leq \beta \leq q)$ are decomposed to the subsets $\alpha_1, \ldots, \alpha_s$ $(\beta_1, \ldots, \beta_t)$ with the same values $u_{\alpha j}$ $(v_{\beta j})$, $1 \leq j \leq r$.